\documentstyle[times,pramana,epsf,floats]{ias}
\begin{document}
\mark{{Direct CP Violation in Neutral Kaon Decays}{W. Wi\'slicki}}
\title{Direct CP Violation in Neutral Kaon Decays}

\author{Wojciech Wi\'slicki, \\
on behalf of the NA48 Collaboration at CERN}
\vspace{3mm}
\address{A. So\l tan Institute for Nuclear Studies, Ho\.za 69, PL-00-681 Warszawa}
\abstract{The final result is presented of the NA48 experiment, performed at the CERN SPS neutral kaon beams, on the direct CP violation parameter Re$(\varepsilon'/\varepsilon)$, as measured from the decay rates of neutral kaons into two pions.
The data collected in years 1997-2001 yield the evidence for the direct CP violation with Re$(\varepsilon'/\varepsilon)=(14.7\pm 2.2)\times 10^{-4}$. 
Description of experimental method and systematics, comparison with the corresponding FNAL result and discussion of some implications for theory are given. 
}

\maketitle
\section{Introduction}
CP violation can occur either via the mixing of CP eigenstates, called {\it indirect} CP violation, or through the interference of amplitudes with different isospins, called {\it direct} CP violation. 
Both processes can be measured in decays of neutral kaons into pions.
In one-loop approximation, the first process is described by box diagrams and parametrized by precisely measured parameter $\varepsilon=(2.28\pm 0.01)\times 10^{-3}$ \cite{hagiwara}, whereas direct CP violation is described by penguin diagrams and quantified using parameter $\varepsilon'$. 
Direct CP violation is an elusive effect which evaded conclusive measurements in experiments of previous generation performed about a decade ago \cite{oldcp}.
On the other hand, the standard model of electro-weak interactions does not provide with the value of Re$(\varepsilon'/\varepsilon)$ precise enough to be comparable with experiments. 
Therefore, it became tempting to perform more accurate measurement with high statistics and well controlled systematics in order to conclude on the existence of this effect.
Two experiments have recently provided results on the measurement of Re$(\varepsilon'/\varepsilon)$: the NA48 at CERN \cite{batley1} and the KTeV at FNAL \cite{alavi1}.

\section{Experimental method}
Experimentally, it is convenient to measure the double ratio $R$ of frequencies of four simultaneously registered decay modes of neutral kaons into two pions, related to the $\varepsilon'$
\begin{eqnarray}
R & = & \frac{|\langle\pi^0\pi^0|K_L\rangle/\langle\pi^0\pi^0|K_S\rangle|^2}{|\langle\pi^+\pi^-|K_L\rangle/\langle\pi^+\pi^-|K_S\rangle|^2} \nonumber \\
  & \simeq & 1-6\,{\mbox Re}(\varepsilon'/\varepsilon)
\label{eq1}
\end{eqnarray}
thus exploiting cancellations of systematic effects contributing in the same way to all processes.

\begin{figure}[htbp]
\begin{tabular}{cc}
\epsfxsize=65mm \epsfysize=50mm \epsfbox{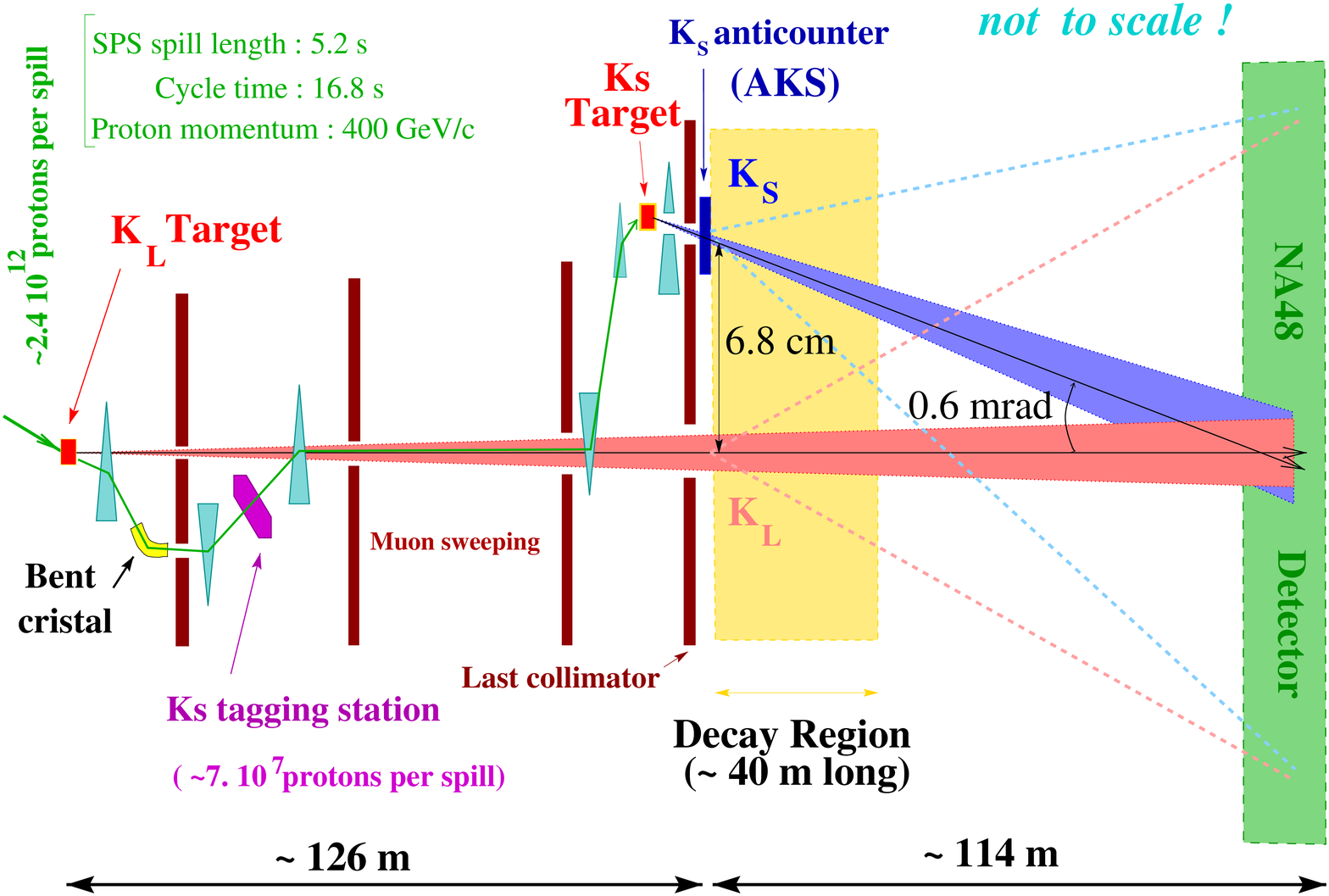} & \epsfxsize=65mm \epsfysize=50mm \epsfbox{detector.eps}
\end{tabular}
\caption{Schematic view (not to scale) of CERN neutral kaon beamline (left) and the NA48 detector (right), located downstream of the decay region.}
\label{fig1}
\end{figure}

In NA48 experiment at CERN, decays of $K_{L,S}$ from two collinear beams, produced by 400 GeV protons on two targets separated by 120 m (fig.\ref{fig1} left), are observed in the same decay region and registered in detectors downstream (fig.\ref{fig1} right).
$K_S$ decays are distinguished from $K_L$ by a coincidence between the decay time and times of protons producing the $K_S$ beam ($\pi^0\pi^0$ mode) or by decay vertex position transverse to beams ($\pi^+\pi^-$ mode).
In order to minimize difference in acceptance due to different lifetimes, $K_L$ decays are weighted as a function of their proper times thus making their spectra similar to those of $K_S$.

Decays $K^0\rightarrow\pi^+\pi^-$ are measured by a magnetic spectrometer composed of four drift chambers with a dipole magnet. 
A liquid crypton electromagnetic calorimeter (LKr), being a hodoscope of $128\times 128$ separately read out cells, is used to reconstruct $K^0\rightarrow\pi^0\pi^0$ decays.

Triggers are designed to reduce the rate of 400 kHz of particles in the spectrometer to less than 10 kHz.

The trigger for $K^0\rightarrow\pi^0\pi^0$ decays operates on analogue sums of $2\times 8$ cells, horizontal and vertical, converted into kinematic variables using look-up tables.
It requires an electro-magnetic energy deposit greater than 50 GeV, a center of energy at calorimeter impact smaller than 25 cm, vertex position not exceeding 5 $K_S$ lifetimes and less than 6 peaks in each of transverse projections.

Charged trigger has two levels.
At the first level it requires coincidence of three fast signals:  coincidence in opposite quadrants of the charged trigger hodoscope, chamber hit multiplicity and total calorimetric energy deposit above 35 GeV.
The second level consists of a fast event-building asynchronous processors reconstructing tracks.

Triggers are accepted if a set of requirements for tracks, vertex and reconstructed mass quality are fulfilled.

\section{Event reconstruction, selections, background rejection and subtracton, acceptance corrections and systematic effects}

For $K^0\rightarrow\pi^0\pi^0$ decays only information from 3--100 GeV photon showers in LKr is used. 
Longitudinal vertex position $z_{vx}$ is reconstructed using shower energies and their impact coordinates at LKr.
The average resolution on $z_{vx}$ is 55 cm and on the kaon energy is 0.5\%.
Invariant masses of $\gamma$-pairs are then computed.
The $\gamma$ pairings are ascribed to $\pi^0$s by requiring that the sum of pair masses is closest to double $\pi^0$ mass and their difference to zero.
The $K^0$ mass is reconstructed from $\pi^0\pi^0$ decays with resolution well below 1 MeV.
Event time is estimated from the most energetic cells in each cluster and is known with accuracy of 220 ps.

\begin{figure}[htbp]                                                                                          
\begin{tabular}{cc}                                                                                           
\epsfxsize=65mm \epsfysize=60mm \epsfbox{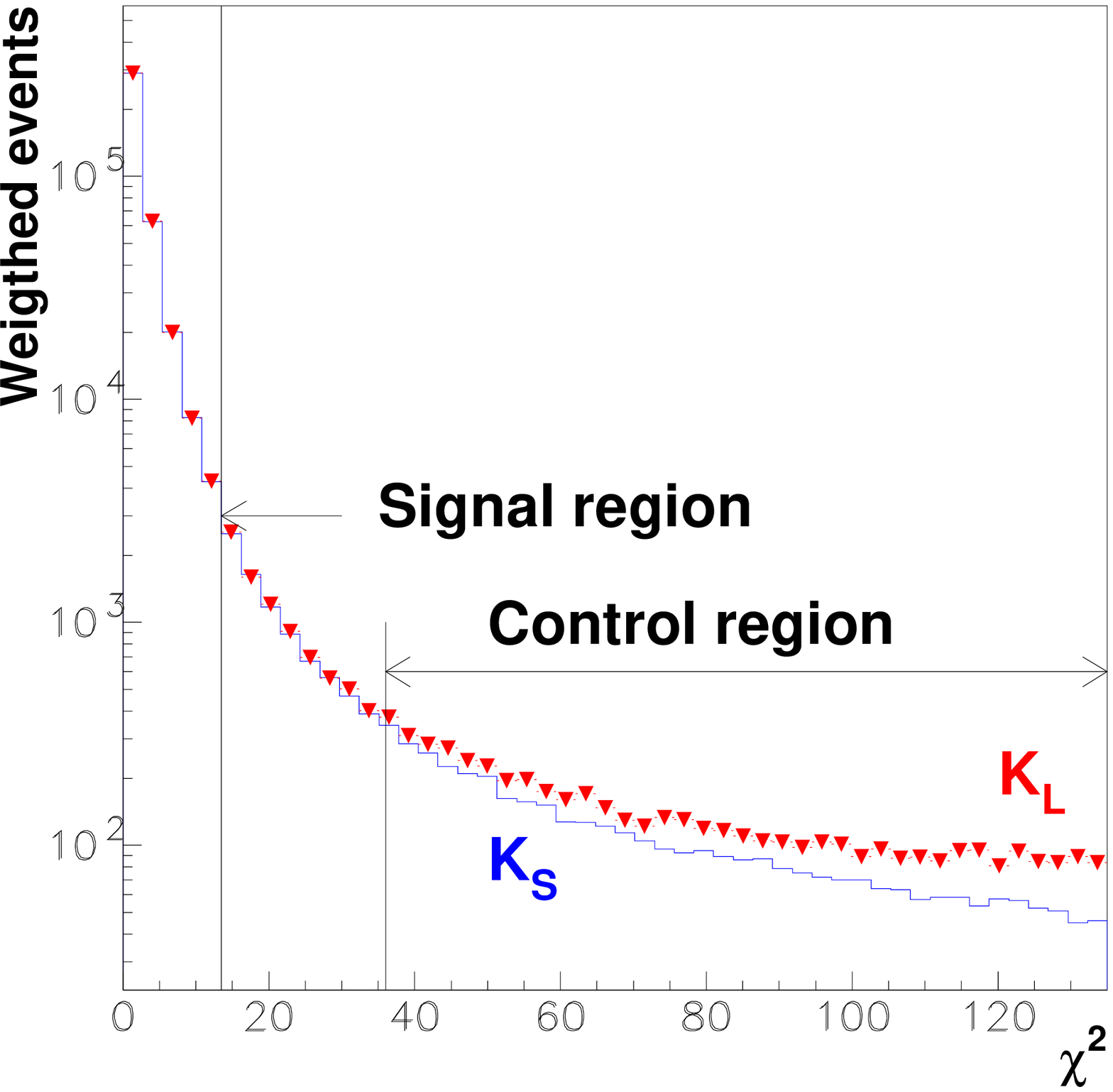} & \epsfxsize=65mm \epsfysize=60mm \epsfbox{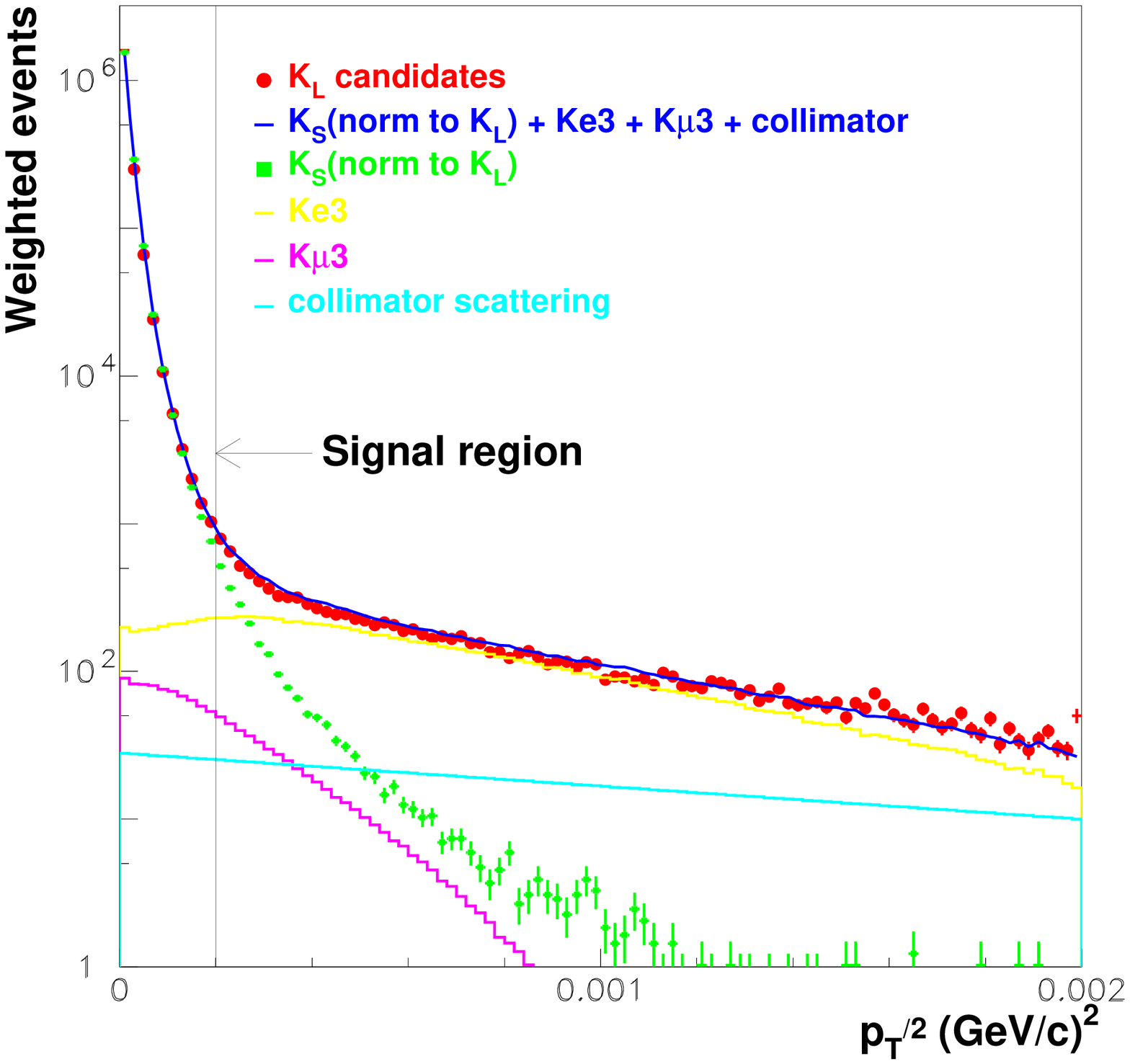}
\end{tabular}
\caption{The $\chi^2$ for $K_{L,S}\rightarrow 2\pi^0$ candidates, where excess in $K_L$ sample for large $\chi^2$ is due to $3\pi^0$ background (left). Comparison of $p_T^2$ for different backgrounds to $K_L\rightarrow\pi^+\pi^-$ (right)}
\label{fig2}
\end{figure}

The background in neutral mode comes uniquely from $K_L\rightarrow 3\pi^0$ decays and is largely suppressed by requiring no additional showers within $\pm 3$ ns around event time.
Remaining background is removed by a constrained for $\chi^2$ built from masses of $\gamma$ pairs (cf. fig.\ref{fig2} left).

The $K^0\rightarrow\pi^+\pi^-$ decays are reconstructed from tracks using hits in the drift chambers, the magnetic field map and detectors alignment.
Average $z_{vx}$ resolution is 50 cm and that for $x_{vx},y_{vx}$ is 2 mm.
The kaon mass is reconstructed in this mode with accuracy of 2.5 MeV.
Event time is given by charged hodoscope with accuracy of 300 ns.

Charged background to $K_S$ decays comes from $\Lambda\rightarrow p\pi^-$ and is suppressed using events with close momenta of the positive and negative particles.
For $K_L$ it comes from semileptonic decays and is rejected by cut on $E/p<0.8$, no hits in muon veto, constraint on effective mass and cut on $p_T$ (fig.\ref{fig2} right).

For the $\pi^+\pi^-$ mode $K_L$ and $K_S$ are distinguished by transverse vertex position.
For the $\pi^0\pi^0$ mode, however, only $z_{vx}$ is reconstructed and one needs time information from the tagger (cf. fig.\ref{fig1} left) and require its coincidence, within $\pm 2$ ns, with event time, thus distinguishing $K_S$ from $K_L$.
Systematic errors related to tagging come from tagging inefficiency (small probability not to register time coincidence) or mistagging (accidental coincdence between proton and event time). 
Both effects are estimated using vertex identification from charged mode and included to systematic error.

Since the acceptance of the apparatus depends on $z_{vx}$ and the $K_L$ and $K_S$ lifetimes are different, a correction is needed to account for difference in acceptances between the $K_L$ and $K_S$.
Distributions of relevant variables for $K_L$ are thus weighted using lifetimes and accounting for interference term.
The procedure was checked and systematic error was calculated using large statistics Monte Carlo ($4\times 10^8$ decays per mode).

Other systematic effects contributing to the value and errors of $R$ include the collimator scatering on the $K_S$ beam, errors in the energy and distance scales from the LKr (inhomogeneities, energy leakage from clusters, non-linearity of energy response, non-Gaussian tails in energy response) and accidental effects due to intensity time dependence and intensity difference between $K_L$ and $K_S$ beams .

\section{The result and discussion}

Overall statistics of reconstructed and accepted events used for determination of $R$ amounts to 65 million (cf. the table).
\begin{center}
\begin{tabular}{||c|c|c|c||} \hline \hline
 \mbox{mode} &  events$\,(\times 10^{-6})$ & \mbox{mode} &  events$\,(\times 10^{-6})$ \\ \hline
$K_L\rightarrow\pi^0\pi^0$ & $4.7$ & $K_L\rightarrow\pi^+\pi^-$ & $21.6$ \\
$K_S\rightarrow\pi^0\pi^0$ & $7.4$ & $K_S\rightarrow\pi^+\pi^-$ & $31.8$ \\ \hline \hline
\end{tabular}
\end{center}
\begin{figure}[htbp]
\begin{tabular}{cc}
\epsfxsize=65mm \epsfysize=70mm \epsfbox{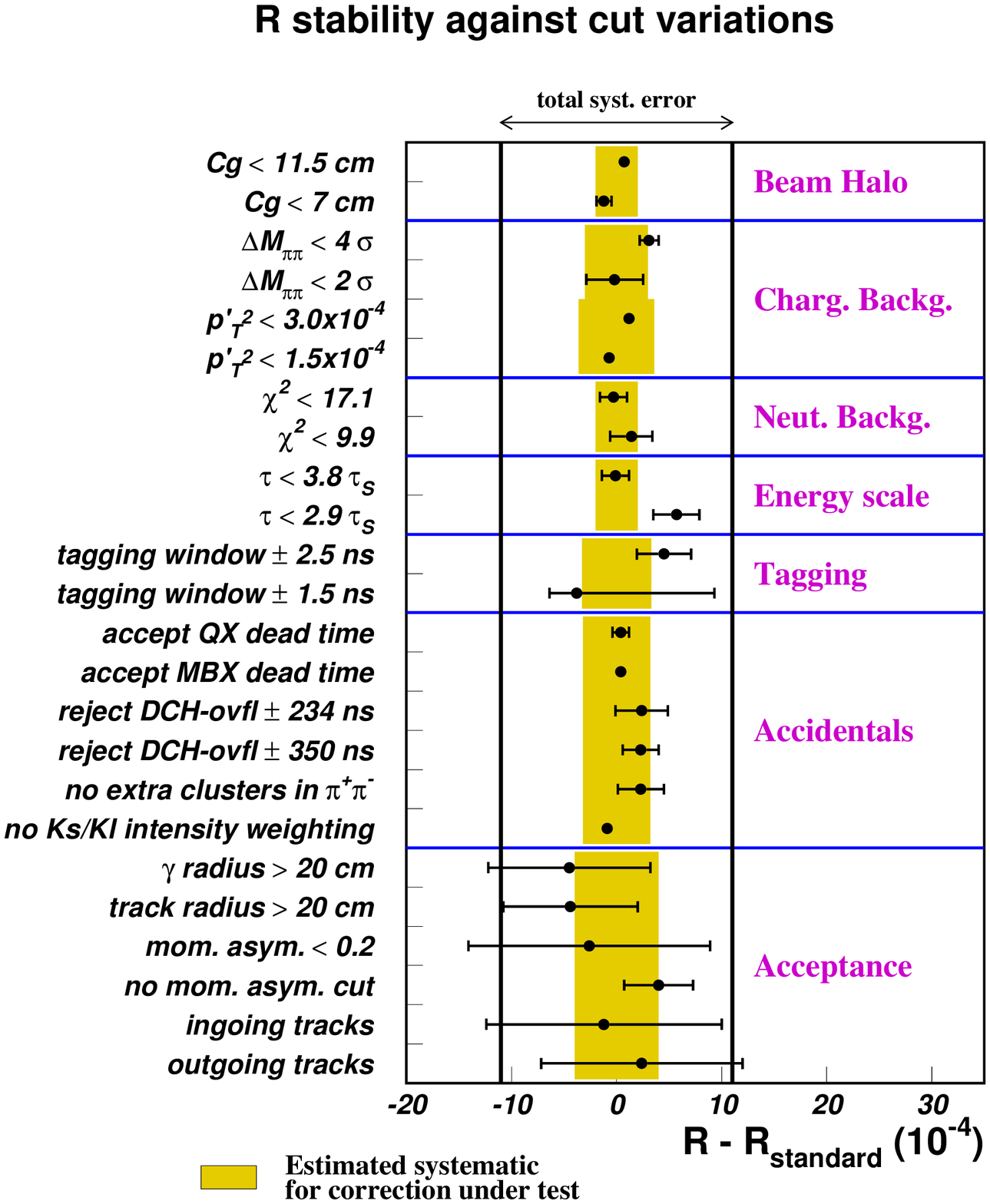} & \epsfxsize=65mm \epsfysize=70mm \epsfbox{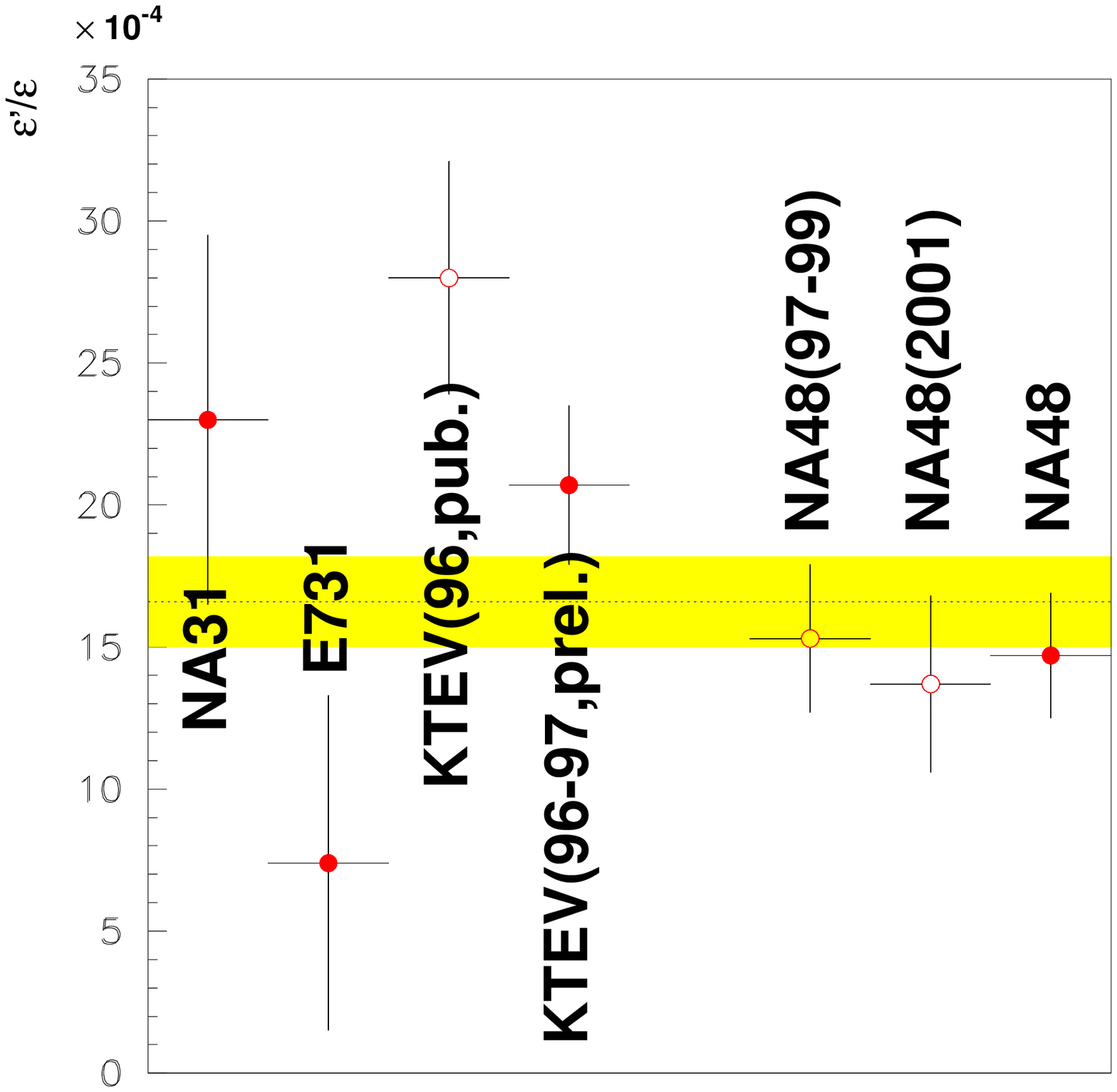}
\end{tabular}
\caption{The summary of corrections and systematic errors to $R$, where $R_{standard}$ corresponds to the final, corrected value of the double ratio, the belt represents systematic error and points with error bars are contributions to the correction on $R$ (left) and world data on the Re$(\varepsilon'/\varepsilon)$ (right).} 
\label{fig3}
\end{figure}
Fig.~\ref{fig3} left summarizes all sources of systematic errors and corrections to R. 
It is seen that major systematic effects to $R$ come from the acceptance and charge background.

The final value of the CP violation parameter obtained by the NA48 Collaboration is
\begin{eqnarray}
\mbox{Re}(\varepsilon'/\varepsilon) & = & (14.7\pm 1.4\pm 0.9\pm 1.5)\times 10^{-4} \nonumber \\
                                    & = & (14.7\pm 2.2)\times 10^{-4}
\end{eqnarray}
where the first error is pure statistical from the $2\pi$ samples, the second is systematic from the statistics of control sample used for its estimation and the third is the contribution of all other systematic errors.
Stability of Re$(\varepsilon'/\varepsilon)$ as a function of the $K^0$ energy was carefully checked and no energy dependence was found.
Also, NA48 finds good consistency of results monitored year by year between 1997 and 2001. 

World data for Re$(\varepsilon'/\varepsilon)$ is presented in fig.~\ref{fig3} right together with the grand average from four exeriments \cite{oldcp,batley1,alavi1} Re$(\varepsilon'/\varepsilon)=(16.7\pm 1.6)\times 10^{-4}$.
Good agreement between experiments can be judged from the figure.
Existence of the direct CP violation in $K^0$ decays is thus proved, providing with fairly precise number to the theory.
It practically excludes Wolfenstein's superweak hypothesis \cite{wolfenstein} which assumed $\varepsilon'=0$.

As commonly known, expanding $\varepsilon'/\varepsilon$ in terms of the matrix elements of the four-quark operators $Q_n$ ($n=1,\ldots,10$), being components of the effective Hamiltonian, one separates it into $\Delta I=1/2$ and $\Delta I=3/2$ parts \cite{branco}.
The first part is dominated by gluonic penguin process and the second one by the $Z^0$ penguin.
The latter is proportional to the square of the top quark mass and the degree of cancellation between the two, due to their destructive interference, is sensitive to it.
Since the top quark mass is already known at present with accuracy of 5 GeV/c$^2$ \cite{hagiwara} and penguin cancellations are known to be large with good precision, larger theoretical uncertainty comes from the long-range part of matrix elements due to strong interactions which have to be calculated non-perturbatively.
Therefore current experimental precission for $\varepsilon'/\varepsilon$ is not really useful for eventual improvement of the CKM matrix parameter $\eta$ until our ignorance about strong parts obscures the whole picture.
Efforts towards estimation of QCD factors from lattice calculations still do not lead to satisfactory results \cite{soni1}, mainly because of usage of uncontrolled approximations, as quenched approximation, in order to make the problem computationally tractable.
For example, currently obtained values of $\varepsilon'/\varepsilon$ are negative and consistent with zero within errors, where theoretical errors are larger than experimental \cite{lattice}.
In addition, contribution to $\varepsilon'/\varepsilon$ from final state interactions is not estimated and it may not be small due to relatively large relative momenta of pions in the final state.

\end{document}